\documentclass{ptapap}

\author{Rados\l{}aw Smolec}[CAMK]
\affil[CAMK]{Nicolaus Copernicus Astronomical Center\\
  Bartycka 18, 00--716 Warszawa, Poland}

\title{The Blazhko Effect}

\begin{document}

\maketitle

\begin{abstract}
We review the recent progress in our understanding of the Blazhko effect in RR~Lyrae stars.
\end{abstract}

\section{Introduction}
%%%%%%%%%%%%%%%%%%%%%%

RR~Lyrae stars are low-mass, typically metal poor, Population II stars. In the HR diagram, they are located within the so-called classical instability strip. Their apparent large-amplitude brightness changes are due to oscillation in the radial modes: radial fundamental mode (RRab stars) or radial first overtone mode (RRc stars). Simultaneous pulsation in these two radial modes is also possible (RRd stars). As excellent distance indicators, RR~Lyrae stars are invaluable in studies of the Galactic structure and evolution.

RR~Lyrae stars are considered a textbook examples of simple, single-periodic radially pulsating stars. In fact, they are much more complex and display phenomena that are not well understood. Recent analyses of top-quality ground-based and space-based photometry revealed that additional low-amplitude variability, attributed to nonradial modes, may be common, at least in RRc stars -- see e.g. \cite{szabo_corot,pamsm15,nsd15}; for recent review, see \cite{pam14}. Another puzzling effect, the Blazhko modulation, was discovered more than hundred years ago \citep{Blazhko}. It manifests as a quasi-periodic modulation of pulsation amplitude and/or phase. Its cause remains a mystery. These proceedings are focused on recent progress in our understanding of the Blazhko phenomenon. 

\section{Blazhko effect: new discoveries from the ground-based and space-based observations}

Blazhko effect is one of the most stubborn problems of stellar astrophysics. Many efforts are devoted to study the phenomenon, both on observational side and on theoretical side. Until recently, the observations were limited to the ground-based only. RR~Lyr, eponym of the class and a Blazhko variable, was a target of multi-site campaign \citep{rrl_multisite}. Many Blazhko variables were discovered in the data gathered by the photometric sky surveys such as OGLE \citep[e.g.,][]{mizerski,o3_rrl} or MACHO \citep[e.g.][]{macho}. Excellent observations, allowing detailed study of the light curve changes during the Blazhko cycle, were collected by the Konkoly Blazhko Survey \citep[e.g.,][]{kbs_mwlyr,sodorCZLac}. These observations also revealed that the Blazhko phenomenon might be common, at least in RRab stars, as it was detected in nearly 50\thinspace \% of RRab stars for which high-quality photometry was gathered \citep{kbs_frequencyrate}. 

Recently, with the advent of space telescopes, {\it CoRoT} and {\it Kepler}, our knowledge about the Blazhko effect significantly increased. Nearly continuous and long-term monitoring of few tens of RR~Lyrae stars revolutionized our knowledge about the Blazhko effect. {\it CoRoT} observed 6 Blazhko variables, {\it Kepler}, in its original Cygnus field monitored for 4 years, observed 18 Blazhko stars (not all stars were observed for 4 years). {\it CoRoT} is no longer functional. {\it Kepler's} capabilities are limited, after a failure of its two reaction wheels. It continues the mission as {\it K2}, and has observed tens of RR~Lyrae stars, but these observations are limited to 90\thinspace days and are of slightly lower precision than in the original mission \citep[initial results were published by][]{molnar}. The 24 Blazhko stars observed by {\it CoRoT} and by {\it Kepler} in its original field, will serve as a benchmark for Blazhko studies in the coming years. In Tab.~\ref{tab:bl} we summarise the basic properties of these stars together with references to the original papers. The most important findings are summarised below. The progress on theoretical side is reported in Sect.~\ref{sec:models}.

\begin{table}[t!]
\setlength{\tabcolsep}{5pt}
\begin{tabular}{rrrrrr}
Star's id  & Other id  & $P_{\rm F}$\thinspace (d) & $P_{\rm BL}$\thinspace (d) & Remarks   & Ref. \\
\hline
\multicolumn{6}{l}{\it CoRoT targets:}\\
0100689962 & V1127~Aql & $0.35600$ & $26.88$                 & PD,2O        & 1,5 \\
0100881648 &           & $0.60719$ & $59.77$                 &              & 5 \\
0101128793 &           & $0.47193$ & $17.86$                 & PD,2O        & 2,5 \\
0101503544 &           & $0.60509$ & $25.60$                 & PD,(nr)      & 5 \\
0103922434 &  V922~Oph & $0.54138$ & $54.5$                  & PD,2O        & 5 \\
0105288363 &           & $0.56744$ & $35.6$                  & 2O           & 3,4,5 \\
\hline
\multicolumn{6}{l}{\it Kepler targets:}\\
   3864443 & V2178~Cyg & $0.48695$ & $213$, $(167.5)$        & 2O,(PD)      & 8,13 \\ 
   4484128 &  V808~Cyg & $0.54786$ & $92.16$, $(\sim\!1000)$ & PD,(2O)      & 7,8,13 \\
   5559631 &  V783~Cyg & $0.62070$ & $27.6667$               &              & 6,8,13 \\ 
   6183128 &  V354~Lyr & $0.56169$ & $849$                   & 2O,(PD),(nr) & 8,13 \\ 
   6186029 &  V445~Lyr & $0.51309$ & $54.83$, $146.9$        & PD,1O,2O     & 8,10,13 \\ 
   7021124 &           & $0.62248$ & $(1400)$                & 2O           & 15 \\
   7198959 &    RR~Lyr & $0.56684$ & $39.1$                  & PD,1O        & 6,7,8,9,11,14 \\ 
   7257008 &           & $0.51179$ & $39.67$, $(>900)$       & PD,2O        & 12,13 \\
   7505345 &  V355~Lyr & $0.47370$ & $31.02$, $16.24$        & PD,2O        & 7,8,13 \\
   7671081 &  V450~Lyr & $0.50462$ & $123.8$, $80.5$         & 2O           & 8,13 \\
   9001926 &  V353~Lyr & $0.55680$ & $71.8$, $132.2$         &              & 8,13 \\
   9508655 &  V350~Lyr & $0.59424$ & $30.6$, $56.4$          & 2O           & 15  \\
   9578833 &  V366~Lyr & $0.52703$ & $62.84$, $29.29$        & (2O)         & 8,13 \\
   9697825 &  V360~Lyr & $0.55758$ & $52.03$, $21.07$        & 2O,(PD)      & 8,13 \\
   9973633 &           & $0.51078$ & $67.2$, $27.17$         & PD,2O        & 12,13 \\
  10789273 &  V838~Cyg & $0.48028$ & $59.7$                  & (2O),(PD)    & 12,13 \\
  11125706 &           & $0.61322$ & $40.21$, $58.9$         &              & 8,13 \\
  12155928 & V1104~Cyg & $0.43639$ & $52.02$                 &              & 8,13 \\
\hline
\end{tabular}
\caption{Basic properties of Blazhko RRab stars observed by {\it CoRoT} and by {\it Kepler} in its original Cygnus field. Star's id, pulsation period and modulation period(s) are given. {\bf Remarks:} `PD' -- period doubling detected, `1O', `2O' -- additional radial modes detected, `nr' -- additional nonradial modes detected; `()' indicate marginal detection. {\bf References:} 
1 -- \cite{chadid_v1127Aql}, 
2 -- \cite{poretti_corot}, 
3 -- \cite{gugg_corot}, 
4 -- \cite{chadid_gugg}, 
5 -- \cite{szabo_corot}, 
6 -- \cite{kol10}, 
7 -- \cite{szabo_pd},
8 -- \cite{benko10},
9 -- \cite{kol11}, 
10 -- \cite{gugg_kepler}, 
11 -- \cite{molnar_rrl},
12-- \cite{nemec13},
13 -- \cite{benko14},
14 -- \cite{LeBorgne},
15 -- \cite{bs15}.}
\label{tab:bl}
\end{table}

{\bf Incidence rate of the Blazhko phenomenon.} The incidence rate is certainly very high, at least for RRab stars. Top-quality ground-based observations and observations of the space telescopes indicate, that the incidence rate is around 50\thinspace\%. The reader is referred to the in-depth analysis by \cite{Kovacs_review}, in particular his tab.~1. The data are much more scarce for RRc stars; these stars were not extensively observed from space yet. Incidence rates are clearly smaller; those reported in the analyses of ground-based data are below 10\thinspace \% \citep{nagy,mizerski}.

\begin{figure}
\centering
\includegraphics[width=0.7\textwidth]{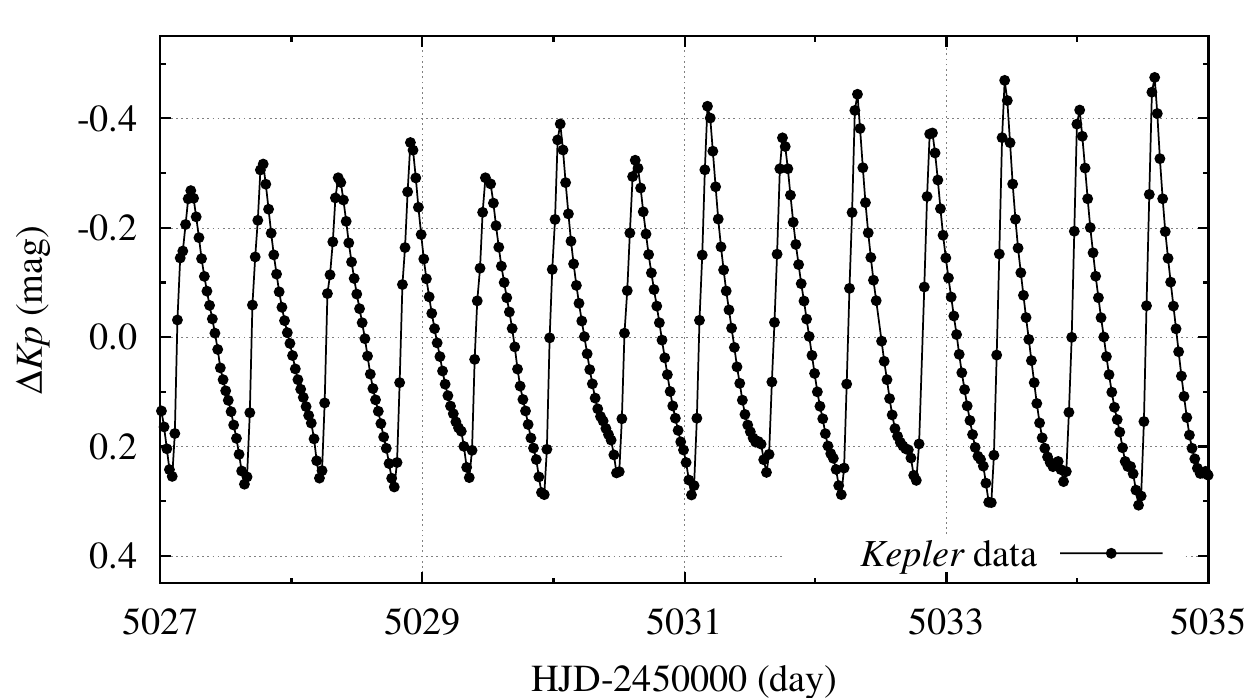}
\caption{Illustration of period doubling in the light curve of RR~Lyr observed with {\it Kepler}.}
\label{fig:pd}
\end{figure}

{\bf Period doubling effect.} This is the most intriguing phenomenon detected thanks to space observations. It is illustrated in Fig.~\ref{fig:pd} with the help of {\it Kepler} data for RR~Lyr, in which the effect was discovered first \citep{kol10}. It manifests as alternating maxima and minima of the pulsation cycles in the light curve. In the frequency spectrum, signals close to half-integer frequencies, i.e. close to $(2n+1)/2\nu_{\rm F}$ ($n\!=\!0,1,2,\ldots$; $\nu_{\rm F}$ -- frequency of the fundamental mode), are detected. In-depth study was done by \cite{szabo_pd}. Detailed analysis of the {\it CoRoT} and {\it Kepler} data \citep{benko14,szabo_corot} revealed the phenomenon in 14 stars i.e. in $\sim$58\thinspace\% of Blazhko stars observed from space. These stars are marked with `PD' in Tab.~\ref{tab:bl}. We note that sometimes the effect is very weak. In some cases however, amplitude of the alternations may be as large as $0.1$\thinspace mag; the effect should be easily detected from the ground. It was not, due to unfortunate pulsation period of RR~Lyrae stars, typically around $0.5$\thinspace d, which prevents observation of consecutive pulsation cycles from (typically a single main site on) Earth.

The term `period doubling' may be misleading in the context of Blazhko variables. Period doubling effect is well known dynamical phenomenon that occurs in type-II Cepheids \citep[e.g.,][]{o3_blg_t2cep,smolecPDBLHer}. In these stars, the alternations are permanent (although pulsation, in particular of RV~Tau stars, may be irregular) and hydrodynamic models indicate that they may represent the first step in the period doubling route to chaos \citep[e.g.][]{kb88}. In the frequency spectrum, additional signals are centered exactly at the half-integer frequencies. This is not the case for Blazhko RR~Lyrae stars. The effect is observed only at some phases of the modulation cycle; its nature is {\it intermittent}. In the frequency spectrum, the additional signals are not centered at $(2n+1)/2\nu_{\rm F}$, but significant offset is sometimes present, as already reported by \cite{szabo_pd} or \cite{benko14} and also analysed by \cite{bryant_pd}. For these reasons, \cite{Kovacs_review} suggests to call the effect `amplitude alternation' rather than period doubling. We note that \cite{szabo_pd} conducted some simulations and showed, that the intermittent nature of the phenomenon may lead to the appearance of complex structures in the frequency spectrum close to half-integer frequencies. In these structures, the highest peaks are not necessarily centered at $(2n+1)/2\nu_{\rm F}$.

With the help of hydrodynamic modelling, \cite{kms11} were able to produce the period doubling effect in non-modulated RR~Lyrae models \citep[so far the Blazhko-like modulation was not reproduced in models of RR~Lyrae stars,][]{smolec_epj} and traced its origin to the 9:2 resonance between the fundamental mode and the ninth overtone (which is a surface mode, trapped in the outer model layers). This result, and lack of the phenomenon in the non-modulated RR~Lyrae stars \citep{nemec11,szabo_corot}, motivated a new model behind the Blazhko modulation \citep[][see next Section]{bk11}.

{\bf Additional modes: radial and nonradial.} A low noise level in the space-based observations allows one to detect periodicities of very low amplitude, even well below 1\thinspace mmag. Indeed, the frequency spectra of Blazhko variables observed from space contain many additional signals beyond the fundamental mode frequency and its harmonics, $k\nu_{\rm F}$, and multiplet structures centered at $k\nu_{\rm F}$. In the low-frequency range a signal corresponding to the modulation frequency (and its harmonics) is often detected. Signals located close to $(2n+1)/2\nu_{\rm F}$ are attributed to period doubling phenomenon. Instrumental frequencies and their combinations are also common in the data, hence, careful analysis of signals at unexpected locations is always needed. Other signals detected in the frequency spectrum fall into two categories: those that can be attributed to radial modes (to the first or to the second overtone) and those that can not. In Tab.~\ref{tab:bl} stars with such signals are marked with `1O'/`2O' or with `nr', respectively. Additional radial modes are recognized based on characteristic period ratios. In particular, for the second overtone, which is detected in Blazhko variables very often (17 stars in Tab.~\ref{tab:bl}, $\sim\!70$\thinspace\% of the sample), we have $P_{\rm 2O}/P_{\rm F}\!\approx0.58\!-\!0.595$. In several stars additional periodicities with $P_{\rm x}/P_{\rm F}$ around $0.7$ were detected and initially identified as corresponding to first overtone or to nonradial modes. \cite{bsiaus301} showed, that in the majority of cases, these frequencies can be interpreted as linear combinations of radial mode frequencies (F and 2O) and radial mode and half-integer frequencies. This is the case for e.g., V360~Lyr and V354~Lyr \citep[1O detection claimed in][]{benko10}, and V1127~Aql \citep[nonradial mode detection claimed in][]{chadid_v1127Aql,szabo_corot}, in which additional signals in the frequency spectrum can be interpreted as linear combination $2(\nu_{\rm 2O}\!-\!\nu_{\rm F})$ \citep{bsiaus301,benko14}. Detection of the first overtone is claimed in only 2 stars. In other two stars significant unidentified frequencies were reported (marked with `nr' in Tab.~\ref{tab:bl}). We stress that amplitude of the additional signals (including 2O) is always very low and sometimes complex structures are present in the frequency spectrum \citep[close double, or multiple peaks, see fig.~6 in][]{benko14}.   

Two stars deserve more attention. V350~Lyr and KIC7021124 were first regarded as non-modulated stars \citep{benko10,nemec11}. Presence of the second overtone in their frequency spectra, motivated \cite{bs15} to search for the Blazhko modulation. After reanalysis of its photometric data and analysis of the light curves and O$-$C diagrams, the Blazhko effect was indeed detected (these stars are reported in Tab.~\ref{tab:bl}). At the moment, none of the nonmodulated RRab stars observed from space, shows additional periodicities. Hence, \cite{bs15} concluded that additional modes appear only in the presence of the Blazhko effect.

{\bf Modulation patterns. Multiperiodic modulation.} 
Among Blazhko stars a variety of modulation patterns are observed. For nice visualtization of space observations the reader is referred to figs. 4 and 6 in \cite{szabo_corot} and fig.~4 in \cite{benko14}. Modulation periods range from a few days to more than thousand of days. Typically, both pulsation amplitude and pulsation phase are modulated. Modulation amplitudes cover a wide range. Thanks to space observations, tiny modulations were revealed (V350~Lyr and KIC7021124). Modulation may be fairly regular and single-periodic as is the case of V1104~Cyg. Nice examples are also known in the ground-based observations, the reader is referred to on-line gallery at:

\texttt{http://users.camk.edu.pl/smolec/blazhko/} 

\noindent However, with the precision of space photometry, multiperiodic and irregular modulation is rather a rule. Strong irregularities are not rare. Well studied examples are {\it CoRoT} id 105288363 \citep{gugg_corot} or RR~Lyr, in which secondary 4\thinspace yr modulation is postulated [\cite{detre4yr}, see also \cite{kol11}]. At the end of regular {\it Kepler} observations, the Blazhko effect vanished in the star \citep{LeBorgne}. The apparent differences in consecutive modulation cycles may also result from multiperiodic modulation. Two modulations are not rare and are detected both in ground-based observations \citep[e.g.,][B\k{a}kowska \& Smolec, these proceedings]{sodorCZLac,skarka_catalog} and are detected with {\it Kepler} \citep[][Tab.~\ref{tab:bl}]{benko14}. In the space data, secondary modulations were often revealed or confirmed with the help of O$-$C analysis. \cite{benko14} noted that ratio of the modulation periods is often close to the ratio of small integers. Similar example in the ground-based observations is reported e.g. in \cite{sodorCZLac}. Another interesting feature is presence of sub-harmonics of the modulation frequency, i.e. significant peaks at $\nu_{\rm BL}/2$ in the low frequency range, and peaks separated by $\nu_{\rm BL}/2$ at the fundamental mode and its harmonics. It is reported in {\it Kepler} stars \citep{benko14}, in ground-based observations \cite[e.g.][]{sodorCZLac}, and in modulated RRd stars \citep[][Sect.~\ref{sec:rrd}]{SmolecBlRRd}.

\section{Models proposed to explain the Blazhko phenomenon}\label{sec:models}
%%%%%%%%%%%%%%%%%%%%%%%%%%%%%%%%%%%%%%%%%%%%%%%%%%%%%%%%%%%%%%%%%%%%%%%%%%%%%

Several models were proposed to explain the Blazhko phenomenon. Some of them are rather ideas, as they lack rigorous physical description. Testing of the models faces a basic difficulty: we still lack realistic, 3D models that could correctly describe turbulent convection present in the envelopes of RR~Lyrae stars and could describe the nonlinear interaction between the radial modes, and radial and nonradial modes. The amplitude equation formalism, that is often invoked, relies on several parameters (saturation and coupling coefficients) that are very difficult to compute.  Nevertheless, some models could be ruled out on observational and/or theoretical grounds. Below we just provide the list of models with some basic description and refer the reader to original papers for more details. For extensive review and critique of the early models (first three in the list below) the reader is referred to the review by \cite{kovacs_aip}.
%\smallskip

 $\bullet$ Magnetic oblique rotator/pulsator model, proposed by \cite{Shibahashi}. Requires strong dipole magnetic field which is not detected \citep[e.g.][]{kol09}. Also, the model predicts strictly periodic modulation and symmetric quintuplet structures in the frequency spectrum. To the contrary, observed modulation may be strongly irregular. In the frequency spectrum, asymmetric multiplets are observed in stars with the top-quality photometry.
%\smallskip

 $\bullet$ Resonant nonradial rotator/pulsator model, see e.g. \cite{nd01}. Modulation is due to rotational splitting of nonradial $\ell=1$ modes that are in 1:1 resonance with the fundamental mode. The model predicts strictly periodic modulation and symmetric triplets in the frequency spectrum which, as noted above, is not the case.
%\smallskip

 $\bullet$ The idea by \cite{Stothers}: variable magnetic field affects the efficiency of envelope convection and causes the modulation of pulsation. The idea lacks rigorous elaboration. For recent critique of this idea see \cite{Smolec_Stothers} and \cite{Molnar_Stothers}.
%\smallskip

 $\bullet$ 9:2 radial mode resonance model, proposed by \cite{bk11}. According to this model the 9:2 resonance between the fundamental mode and the ninth overtone, which is claimed to be responsible for the period doubling effect \citep{kms11}, may also cause modulation of pulsation. This conclusion is based on the analysis of amplitude equations. The resulting modulation may be quasi-periodic or chaotic. We lack realistic hydrodynamic calculations confirming the hypothesis, although similar mechanism (3:2 resonance) may work in hydrodynamic BL~Her-type models \citep{sm12}. 
%\smallskip

 $\bullet$ Atmospheric shocks' dynamics as a cause of the Blazhko effect -- a model proposed by \cite{Gillet}, based on analysis of monoperiodic, radiative hydrodynamic models. It is not clear how a transient first overtone, crucial in this model, develops in a single-periodic fundamental mode model.
%\smallskip

 $\bullet$ Nonresonant radial-nonradial mode interaction, e.g. \cite{Cox2013}, recently developed by \cite{Bryant_model}. Assumes excitation of two modes, radial and nonradial, of nearly the same frequency, that are not phase-locked. Model based on simple analytical considerations. We lack tools that allow realistic study of radial-nonradial mode interaction in nonlinear regime.

\section{Blazhko effect in double-mode RR~Lyrae stars}\label{sec:rrd}
%%%%%%%%%%%%%%%%%%%%%%%%%%%%%%%%%%%%%%%%%%%%%%%%%%%%%%%%%%%%%%%%%%%%%

\begin{figure}
\centering
\includegraphics[width=0.68\textwidth]{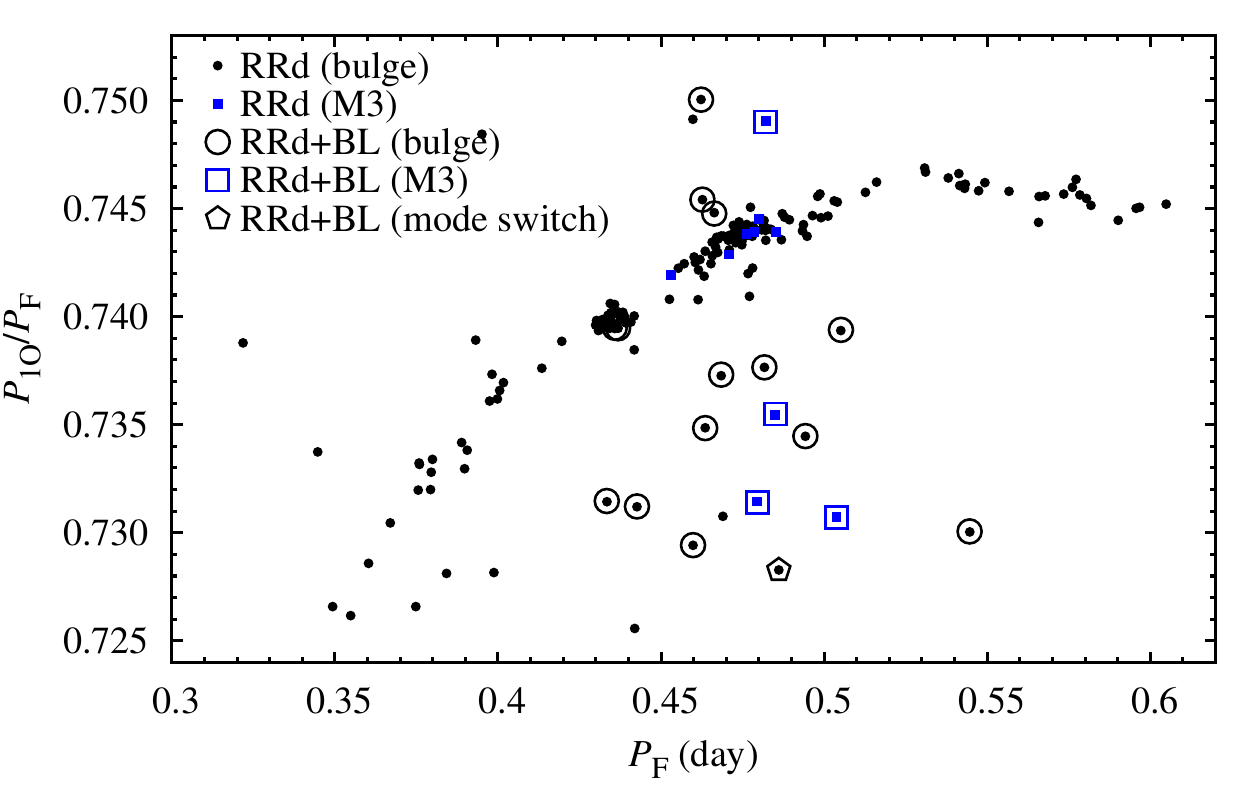}
\caption{Petersen diagram for RRd stars with and without modulation of pulsation.}
\label{fig:pet}
\end{figure}

Recently, the Blazhko effect was discovered in the double-mode RR~Lyrae stars pulsating simultaneously in the radial fundamental and first overtone modes (RRd). Few stars were identified in the OGLE-IV Galactic bulge data by \cite{IgorBlRRd}. Additional stars and detailed analysis of the OGLE sample of Blazhko RRd stars (15 objects) was conducted by \cite{SmolecBlRRd}. Another four variables of the same type were identified by \cite{JurcsikBlRRd} in their observations of the globular cluster M3. The most striking feature of these stars, which we illustrate in the Petersen diagram in Fig.~\ref{fig:pet}, is atypical period ratio. In this plot, the majority of RRd stars form a tight progression; the period ratio increases from $0.725$ to $0.746$ with the increasing pulsation period, up to $P_{\rm F}\approx 0.52$d. Then, a mild decrease of the period ratio with the increasing period is observed. Most of the modulated stars, marked with large open symbols, are beyond this progression. Still, models show that the two dominant periodicities can be easily explained as corresponding to two radial modes \citep{smolec_visegrad}. In some of the stars, only one radial mode is modulated. In other stars, two modes are modulated, but sometimes modulation period is different for the fundamental mode than for the first overtone mode [see tab.~2 in \cite{SmolecBlRRd} and tab.~1 in \cite{JurcsikBlRRd}]. In the OGLE sample, modulation properties are commonly nonstationary. Also pulsation amplitude and phase of the radial modes may strongly vary on a long time scale of a few hundred days. An extreme case is OGLE-BLG-RRLYR-13442, which switched the pulsation mode from RRab to RRd during the OGLE monitoring \citep{IgorBlRRd}. Needless to say, the mechanism behind the Blazhko modulation in RRd stars remains a mystery.

\acknowledgements{This research is supported by the Polish National Science Centre through grant DEC-2012/05/B/ST9/03932. I am grateful to Pawe\l{} Moskalik and Katrien Kolenberg for reading and commenting the manuscript.}

\bibliographystyle{ptapap}
\bibliography{smolec}

\end{document}